\documentclass[
 aps,
 ,twocolumn
]{revtex4}
\usepackage{graphicx}
\usepackage{amsmath}
\usepackage{amssymb}

\begin{document}

\title{Quantum communication in the presence of a horizon}

\author{Daiqin Su}
\author{T.~C.~Ralph}
\affiliation{Centre for  Quantum Computation and Communication Technology, School of Mathematics and Physics,
The University of Queensland, St. Lucia 4072, Queensland, Australia}

\begin{abstract}
Based on homodyne detection, we discuss how the presence of an event horizon affects quantum communication between an inertial partner, Alice, and a uniformly 
accelerated partner, Rob. We show that there exists a low frequency cutoff for Rob's homodyne detector that maximizes the signal to noise ratio and it
approximately corresponds to the Unruh frequency. In addition, the low frequency cutoff which minimizes the conditional variance between Alice's
input state and Rob's output state is also approximately equal to the Unruh frequency. Thus the Unruh frequency provides a natural low frequency cutoff
in order to optimize quantum communication of both classical and quantum information between Alice and Rob.
\end{abstract}

\pacs{03.67.Dd, 42.50.Dv, 89.70.+c}




\maketitle

\vspace{10 mm}

\section{Introduction}
One important task of relativistic quantum information\cite{Peres04} is to investigate how relativistic motion and gravitational fields affect the storage, transfer 
and processing of quantum information. Early works mainly studied global states of quantum fields, for example, the effects of acceleration on the entanglement 
of global states\cite{Milburn03,Fuentes05}. Recently, a general framework for projective measurements on a localized single mode of the quantum field was proposed\cite{Dragan12}. 
As a specific realization of localized projective measurements, homodyne detection was proposed as a way to model efficient, directional quantum communication 
between two localized parties in a relativistic quantum field theory scenario\cite{Downes13}.
An interesting case is the quantum communication between an inertial partner and a uniformly accelerated partner, in which the Unruh effect\cite{Unruh76}
is expected to play an important role. In \cite{Downes13}, an inertial sender, Alice, sends a coherent state signal and a local oscillator to
an accelerated receiver, Rob, who then performs homodyne detection in his own frame. Approximate analytic solutions were obtained in the case
the wave packet sent by Alice is well localized in the right Rindler wedge. In this paper, we generalize this work to the case in which the wave packet
straddles the future horizon of Rob. Similar scenario was considered to study quantum entanglement through the event horizon\cite{Dragan13}.
As a result, Rob can only access part of the signal and local oscillator. Generally, the signal and noise received by Rob are 
divergent if Rob's detector can detect arbitrarily low frequency particles. This is because in the horizon-straddling case Rob can still detect particles at late times 
when his velocity approaches the speed of light, resulting in large redshift of the signal and local oscillator. While, under some special conditions,
the signal and local oscillator received by Rob remain finite no matter what low frequency cutoff he chooses. In order to get finite results generally, 
and to correspond with physical detectors,
we introduce a low frequency cutoff. We find that there exists a low frequency cutoff that 
maximizes the signal to noise ratio. Interestingly, this low frequency cutoff approximately corresponds to the Unruh temperature,
and we thus call it the Unruh frequency. In addition, we calculate the conditional variance and find that the low frequency cutoff that minimizes the conditional variance is
also approximately equal to the Unruh frequency.

For simplicity, we consider the massless scalar field in $(1+1)$-dimensional Minkowski spacetime. Generalization to $(3+1)$-dimensional Minkowski spacetime is straightforward
by making the paraxial approximation and taking into account the expansion of the transverse shape of the wave packet during its propagation.
There are two inequivalent ways to quantize the 
massless scalar field in Minkowski spacetime\cite{Fulling73}, one for inertial observers and the other for uniformly accelerated observers. This results in different
particle concepts for these two sets of observers. In particular, the vacuum state for inertial observers looks like a thermal state when observed by uniformly
accelerated observers, which is known as the Unruh effect\cite{Unruh76}. We introduce Minkowski coordinates $(t,x)$ in the inertial frame and Rindler 
coordinates $(\tau,\xi )$ in the accelerated frame. The transformations between them are
\begin{equation}\label{coordinate transformation}
t = \frac{1}{a}e^{a \xi} \text{sinh}(a \tau), ~~~~ x = \frac{1}{a}e^{a \xi} \text{cosh}(a \tau),
\end{equation}
where $a$ is the proper acceleration of the accelerated observer who travels along the worldline $\xi = 0$ in the right
Rindler wedge. 

The paper is organized as follows: in Sec.\ref{HD:Rob}, we introduce some basic concepts of homodyne detection in an accelerated frame and derive 
general expressions for the expectation value and variance of Rob's output signal. We then calculate the signal to noise ratio and conditional variance 
in the horizon-straddling case for different low frequency cutoffs in Sec.\ref{horizon-straddling}. Finally, we conclude in Sec.\ref{conclusion}.
\begin{figure}[ht!]
\includegraphics[width=5.0cm]{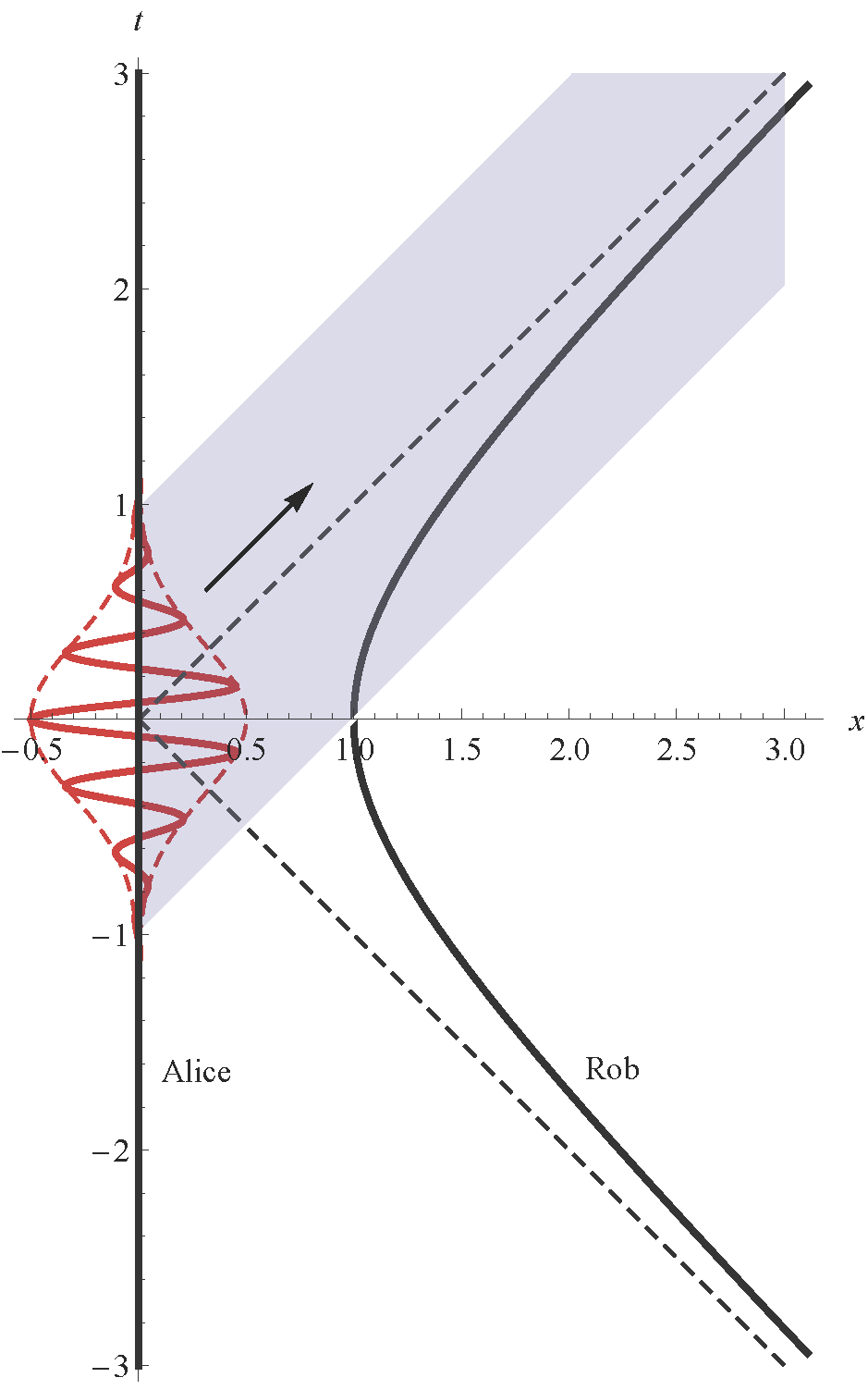}
\caption{\footnotesize (color online). Alice(static) sends Rob(accelerated) a Gaussian wave packet which straddles Rob's future horizon.} 
\label{scenario:Fig}
\end{figure}

\section{Homodyne Detection in an accelerated frame}\label{HD:Rob}
In the inertial frame, the massless scalar field is quantized in the usual way,
\begin{equation}
\hat{\Phi}(t,x) = \int_{0 }^{\infty} d k_s \big(g_{k_s} \hat{a}_{k_s} + g^*_{k_s} \hat{a}^{\dag}_{k_s} \big)+ (\text{left-moving parts}),
\end{equation}
where $g_{k_s} = \frac{1}{\sqrt{4 \pi k_s}}e^{- i k_s(t-x)}$ are positive frequency right-moving Minkowski plane wave mode functions,
$ g^*_{k_s}$ are negative frequency mode functions, and
$\hat{a}_{k_s}$ ($\hat{a}^{\dag}_{k_s}$) are annihilation(creation) operators of single frequency Minkowski modes obeying 
the usual boson commutation relation
\begin{equation}
[\hat{a}_{k_s},\hat{a}^{\dag}_{k'_s}] = \delta(k_s - k'_s).
\end{equation}
In the accelerated frame, $\hat{\Phi}(\tau,\xi)$ can be expanded as 
\begin{equation}
\hat{\Phi}(\tau,\xi) = \int_0^{\infty} d k_d \big(w_{k_d} \hat{b}_{k_d} + w^*_{k_d} \hat{b}^{\dag}_{k_d} \big) + (\text{left-moving parts}),
\end{equation}
where $w_{k_d} = \frac{1}{\sqrt{4 \pi k_d a}}e^{- i k_d a(\tau-\xi)}$ are positive frequency right-moving Rindler plane wave mode functions,
$w^*_{k_d}$ are negative mode functions,
$\hat{b}_{k_d}$ ($\hat{b}^{\dag}_{k_d}$) are annihilation(creation) operators of single frequency Rindler modes obeying 
boson commutation relation
\begin{equation}
[\hat{b}_{k_d},\hat{b}^{\dag}_{k'_d}] = \delta( k_d - k'_d).
\end{equation}
Here $k_d$ is defined as a dimensionless wave number, which is related to the physical frequency $\omega_d$ by $k_d = \omega_d /a$. 

We consider the scenario that a uniformly accelerated observer, Rob, with proper acceleration $a$ travels along $\xi = 0$ in the right 
Rindler wedge and an  inertial observer, Alice, stays at spatial origin $x = 0$, as shown in Fig.\ref{scenario:Fig}. Alice sends a right-moving signal, a coherent state with 
amplitude $\alpha$, and a local oscillator to Rob. The local oscillator is also a coherent state, but with very large amplitude $\beta\in \mathbb{R}$, 
$\beta\gg |\alpha|$. Rob then performs homodyne detection on the signal using the local oscillator as seen in his own reference frame. The homodyne
detector is formed from two identical photodetectors that detect distinct modes $S$ and $L$ after they have been mixed on a beam splitter. 
The photocurrents from the photodetectors are subtracted to give the output signal. As a result the output of Rob's homodyne detector at 
some time $\tau$(as measured in Rob's frame) is represented by the following operator\cite{Ralphbook}:
\begin{equation}
\hat{O}(\tau, \phi ) = \hat{b}_S(\tau) \hat{b}^{\dag}_L(\tau) e^{i \phi}+ \hat{b}^{\dag}_S(\tau) \hat{b}_L(\tau) e^{-i \phi},
\end{equation}
where $\hat{b}_K$($\hat{b}^{\dag}_K$) are boson annihilation(creation) operators with $K = S,L$. The subscripts $S$, $L$ refer 
to the signal and local oscillator modes, respectively. The relative phase $\phi$ determines the quadrature angle detected. 
Here $\hat{b}_K(\tau)$ are temporally and spatially localized single mode annihilation operators in the perspective of Rob. 
They can be constructed from the single frequency Rindler annihilation operators $\hat{b}_{k_d}$,
\begin{equation}
\hat{b}_K(\tau) = \int d k_d f_K(k_d, \tau) \hat{b}_{k_d},
\end{equation}
where $f_K(k_d, \tau)$ is Rob's detector mode function. In an experiment, Rob would integrate the photocurrent from his 
detector over a time long compared to the inverse of the frequency being analyzed(as will be determined by the frequency of 
the local detector). For later convenience, we define the integrated output signal operator $\hat{X}(\phi)$,
\begin{eqnarray}
\hat{X}(\phi) &=& \int d \tau \hat{O}(\tau, \phi ) \nonumber \\
&=& \int d \tau \big[\hat{b}_S(\tau) \hat{b}^{\dag}_L(\tau) e^{i \phi}+ \hat{b}^{\dag}_S(\tau) \hat{b}_L(\tau) e^{-i \phi}\big].
\end{eqnarray} 
The expectation value of the output signal received by Rob is 
\begin{equation} \label{expectation}
X_{\phi} = \langle \hat{X}(\phi ) \rangle,
\end{equation}
and the variance is
\begin{equation} \label{variance}
V_{\phi} = \langle \hat{X}(\phi )^2 \rangle - \langle \hat{X}(\phi ) \rangle^2.
\end{equation}

Alice prepares coherent states(signal and local oscillator) by displacing the Minkowski vacuum $|0 \rangle$ using the displacement operators
$D_K(\gamma) = \text{exp}[\gamma \hat{a}^{\dag}_K - \gamma^* \hat{a}_K]$, with $\gamma = \alpha, \beta$, and
\begin{equation}
\hat{a}_K = \int d k_s f_{D_K}(k_s,t,x) \hat{a}_{k_s},
\end{equation}
where $f_{D_K}(k_s,t,x)$ is a normalized displacement mode function satisfying $\int d k_s |f_{D_K}(k_s,t,x)|^2 = 1$.
Therefore, $\hat{a}_K$ are also temporally and spatially localized annihilation operators in the perspective of Alice. 
The state that Alice prepares can be written in a compact form,
\begin{equation}
|\alpha, \beta, t \rangle = D_S(\alpha ) D_L(\beta ) |0 \rangle.
\end{equation}
The expectation value of the signal becomes
\begin{equation}
X_{\phi} = \langle 0 | D^{\dag}_L(\beta ) D^{\dag}_S(\alpha ) \hat{X}(\phi) D_S(\alpha ) D_L(\beta ) | 0 \rangle.
\end{equation}

In order to explicitly calculate the expectation value and variance of the signal, we need to know the Bogolyubov
transformation between the Rindler modes and Minkowski modes, which are already given by\cite{Takagi&Crispino}
\begin{equation}\label{Bogolyubov transformation}
\hat{b}_{k_d} = \int d k_s (A_{k_d k_s} \hat{a}_{k_s} + B_{k_d k_s} \hat{a}^{\dag}_{k_s}),
\end{equation}
where 
\begin{eqnarray} \label{coefficients}
A_{k_d k_s} = \frac{i e^{\pi k_d/2}}{2 \pi \sqrt{k_d k_s}} \Gamma(1-i k_d) \bigg(\frac{k_s}{a}\bigg)^{i k_d}, \nonumber \\
B_{k_d k_s} = \frac{i e^{- \pi k_d/2}}{2 \pi \sqrt{k_d k_s}} \Gamma(1-i k_d)\bigg(\frac{k_s}{a}\bigg)^{i k_d}
\end{eqnarray}
are the Bogolyubov coefficients for right-moving waves. Taking into account Eq.(\ref{Bogolyubov transformation}), we can find the identity
\begin{eqnarray}\label{identity}
D^{\dag}_K(\gamma ) \hat{b}_K(\tau) D_K(\gamma ) &=& \hat{b}_K(\tau) + \gamma \int d k_d \int d k_s f_K(k_d, \tau ) \nonumber \\
 && \times \big(A_{k_d k_s} f^*_{D_K}(k_s) + B_{k_d k_s} f_{D_K}(k_s) \big) \nonumber \\
 &\equiv & \hat{b}_K(\tau) + \gamma F_K(\tau). 
\end{eqnarray}
The expressions for $X_{\phi}$ and $V_{\phi}$ can be expanded via Eq.(\ref{identity}).

Although the amplitude of the local oscillator sent by Alice is $\beta$, it is not so when viewed by Rob due to Doppler shift
and Rob's inability to access the whole wave packet. The latter effect is more important in the horizon-straddling case. However,
one has to bear in mind that this does not mean the amplitude of the local oscillator must be attenuated. In fact, it sometimes
can be amplified. Homodyne detection only measures the amplitude without caring about the frequency of the mode.
So it is possible that Rob detects a large amount of low frequency particles but the total energy of these particles is still
smaller than the energy of the original wave packet. If Rob performs homodyne detection
without knowing the amplitude of the local oscillator sent by Alice, he has to measure the strength of the local oscillator by adding
the photocurrents of the two photodetectors. We define the strength of the local oscillator as seen by Rob as
\begin{eqnarray} \label{LO strength}
I &=& \int d \tau \langle \hat{b}^{\dag}_L \hat{b}_L \rangle \nonumber \\
&=& \int d \tau \langle 0 | D^{\dag}_L(\beta ) D^{\dag}_S(\alpha ) \hat{b}^{\dag}_L \hat{b}_L D_S(\alpha ) D_L(\beta ) | 0 \rangle.
\end{eqnarray}
Both the expectation value $X_{\phi}$ and variance $V_{\phi}$ of the signal should be normalized by the strength of the local oscillator. 

Since the Bogolyubov transformation (\ref{Bogolyubov transformation}) is a linear transformation, it is obvious that
$\langle 0 | \hat{b}_K | 0 \rangle = \langle 0 | \hat{b}^{\dag}_K | 0 \rangle = 0$. Taking into account the fact 
that $\beta \gg |\alpha|$, we have
\begin{flalign}
&X_{\phi} \approx \beta \alpha e^{i \phi} \int d \tau F_S(\tau) F^*_L(\tau) + \text{c.c.} ,&  \nonumber\\
&V_{\phi} \approx \beta^2 \int d \tau \int d \tau' F^*_L(\tau) F_L(\tau')\langle 0 | \{\hat{b}_S(\tau), \hat{b}^{\dag}_S(\tau') \} | 0 \rangle, &  \nonumber\\
&I \approx \beta^2 \int d \tau F_L(\tau) F^*_L(\tau),&
\end{flalign}
where $\{\hat{A}, \hat{B} \} = \hat{A}\hat{B} + \hat{B}\hat{A}$ represents anticommutation of two operators.
If we further require that the detector mode function for signal and local oscillator are the same and the displacement mode function for
signal and local oscillator are also the same, then $F_S(\tau) = F_L(\tau)$. The normalized output signal becomes
\begin{eqnarray}\label{normalized signal}
\bar{X}_{\phi} = \frac{X_{\phi}}{\sqrt{I}} & \approx & \sqrt{\int d \tau F_L(\tau) F^*_L(\tau)} (\alpha e^{i \phi} + \alpha^* e^{- i \phi} ) \nonumber \\
& \approx & \frac{\sqrt{I}}{\beta} (\alpha e^{i \phi} + \alpha^* e^{- i \phi} ),
\end{eqnarray}
and the normalized variance becomes
\begin{equation}\label{normalized variance}
\bar{V}_{\phi} = \frac{V_{\phi}}{I} \approx \frac{\int d \tau \int d \tau' F^*_L(\tau) F_L(\tau')\langle 0 | \{\hat{b}_S(\tau), \hat{b}^{\dag}_S(\tau') \} | 0 \rangle}
{\int d \tau F_L(\tau) F^*_L(\tau)}.
\end{equation}
In order to proceed, we need to introduce explicit forms for Rob's detector mode function and Alice's displacement mode function. 
The detector mode function can be written as 
\begin{equation}
f_K(k_d, \tau) = e^{-i k_d a \tau} f_K(k_d).
\end{equation}
It is important that the detector mode function should be well localized spatially and temporally; otherwise, its interpretation 
as a detector following a particular spacetime trajectory is compromised. Thus we consider a detector mode function that is very
broad in $k_d$; in particular, we take $f_K(k_d) \approx \sqrt{a/2 \pi}$ for $k_d \geq  k_{\text{cut}} > 0$ and zero otherwise, where $k_{\text{cut}}$ 
is some low frequency cutoff. We will see that if we do not introduce a low frequency cutoff, $\bar{X}_{\phi}$ and $\bar{V}_{\phi}$ may be
divergent. That means if Rob's detector is accurate enough so that it responds to any low frequency particles,
he will detect very large amounts of low frequency particles. However, in practice, there is always some low frequency below which Rob's
detector cannot detect. 

From Fig.\ref{scenario:Fig} we can see that, in the horizon-straddling case, the wave packet overlaps with Rob's whole future
worldline. That is to say, Rob can detect particles even when $\tau \rightarrow + \infty $. Therefore, the integrals over $\tau$ in 
Eqs.(\ref{normalized signal}) and (\ref{normalized variance}) go from
$- \infty$ to $+ \infty$ and we have the simplification $\int d \tau \frac{a}{2\pi} e^{-i (k_d - k'_d) a \tau} \approx \delta(k_d - k'_d)$.

The displacement mode function can be written as
\begin{equation}
f_{D_K}(k_s, t, x) = e^{-i( \omega_s t - k_s x)} f_D(k_s).
\end{equation}
We assume that the displacement mode function is peaked at a large wave number $k_{so}>0$, much larger than the bandwidth $\sigma$, 
although $\sigma$ is also broad on the wavelength scale. Hence we write $k_s = k_{so} + \bar{k}$, where $k_{so}\gg |\bar{k}|$ for the 
region of wave numbers for which the mode function is nonzero. These are typical approximations used for nonrelativistic quantum 
communication systems. Given this, the displacement mode function becomes $e^{-i k_s ( t -  x)} f_D(k_s)$. In particular, we choose
$f_D(k_s)$ as a Gaussian form,
\begin{equation}
f_D(k_s) = \bigg(\frac{1}{2 \pi \sigma^2}\bigg)^{1/4} \text{exp} \bigg\{{-\frac{(k_s - k_{so})^2}{4 \sigma^2}} \bigg\},
\end{equation}
where $k_{so}/\sigma \gg 1$. One term in Eq.(\ref{coefficients}) can be approximated as 
\begin{equation}
\bigg(\frac{k_s}{a}\bigg)^{i k_d} \approx  e^{i k_s (\frac{k_d}{k_{so}})} e^{i k_d [\text{ln}(k_{so}/a) - 1]},
\end{equation} 
and using the identity
\begin{equation*}
|\Gamma(1-i k_d)|^2 = \frac{\pi k_d}{\text{sinh}(\pi k_d)},
\end{equation*}
we have
\begin{eqnarray}
A_{k_d k_s} A^*_{k_d k'_s} \approx \frac{1}{2 \pi k_{so} (1 - e^{- 2 \pi k_d})} e^{i k_s (\frac{k_d}{k_{so}})} e^{-i k'_s (\frac{k_d}{k_{so}})}, \nonumber \\
A_{k_d k_s} B^*_{k_d k'_s} \approx \frac{e^{-\pi k_d}}{2 \pi k_{so} (1 - e^{- 2 \pi k_d})} e^{i k_s (\frac{k_d}{k_{so}})} e^{-i k'_s (\frac{k_d}{k_{so}})}, \nonumber \\
B_{k_d k_s} A^*_{k_d k'_s} \approx \frac{e^{-\pi k_d}}{2 \pi k_{so} (1 - e^{- 2 \pi k_d})} e^{i k_s (\frac{k_d}{k_{so}})} e^{-i k'_s (\frac{k_d}{k_{so}})}, \nonumber \\
B_{k_d k_s} B^*_{k_d k'_s} \approx \frac{e^{-2 \pi k_d}}{2 \pi k_{so} (1 - e^{- 2 \pi k_d})} e^{i k_s (\frac{k_d}{k_{so}})} e^{-i k'_s (\frac{k_d}{k_{so}})}.
\end{eqnarray}
The strength of the local oscillator received by Rob can be calculated as
\begin{flalign}\label{signal-general}
&I = \beta^2 \int d k_d \int d k_s \int d k'_d \int d k'_s \int d \tau f_L(k_d,\tau) f^*_L(k'_d,\tau)& \nonumber \\
& \times \bigg(A_{k_d k_s} f^*_{D_L}(k_s) + B_{k_d k_s} f_{D_L}(k_s) \bigg)\bigg(A^*_{k'_d k'_s} f_{D_L}(k'_s) & \nonumber \\
&  + B^*_{k'_d k'_s} f^*_{D_L}(k'_s) \bigg) & \nonumber \\
&= \beta^2 \int d k_d \int d k_s \int d k'_s \bigg[ A_{k_d k_s} A^*_{k_d k'_s} f^*_{D_L}(k_s)f_{D_L}(k'_s) +  \nonumber \\
&A_{k_d k_s} B^*_{k_d k'_s} f^*_{D_L}(k_s) f^*_{D_L}(k'_s) + B_{k_d k_s} A^*_{k_d k'_s} f_{D_L}(k_s)f_{D_L}(k'_s) & \nonumber \\
& + B_{k_d k_s} B^*_{k_d k'_s}f_{D_L}(k_s)f^*_{D_L}(k'_s) \bigg]& \nonumber \\
& = \beta^2 \sqrt{\frac{2}{\pi}} \frac{\sigma}{k_{so}} \int d k_d \frac{1}{1 - e^{- 2 \pi k_d}} \bigg \{ e^{-2\sigma^2 [k_d + k_{so}(t-x)]^2/k^2_{so}} & \nonumber \\
& + 2 \text{cos}[2 k_{so}(t-x)] e^{-\pi k_d}e^{-\sigma^2 [k_d + k_{so}(t-x)]^2/k^2_{so}}  & \nonumber \\
&\times e^{-\sigma^2 [k_d - k_{so}(t-x)]^2/k^2_{so}} + e^{-2\pi k_d}e^{-2\sigma^2 [k_d - k_{so}(t-x)]^2/k^2_{so}} \bigg\}.
\end{flalign}
Substituting Eq.(\ref{signal-general}) into Eq.(\ref{normalized signal}),
we thus have a general expression for the expectation value of the signal. 

If $t-x<0$ and $|k_{so}(t-x)| \gg k_{so}/\sigma$, then only the first
term in Eq.(\ref{signal-general}) survives. In addition, the Gaussian part of the integrand can be approximated as a delta function, that is,
$\sqrt{\frac{2}{\pi}} \frac{\sigma}{k_{so}} e^{-2\sigma^2 [k_d + k_{so}(t-x)]^2/k^2_{so}} \approx \delta(k_d + k_{so}(t-x))$.
We can recover the analytic expression for the normalized output signal found in \cite{Downes13},
\begin{equation*}
\bar{X}_{\phi} \approx \frac{\alpha e^{i \phi} + \alpha^* e^{- i \phi}}{\sqrt{1 - e^{- 2 \pi k_{so}|t-x|}}}.
\end{equation*}
In this case, Rob can access nearly the whole wave packet because $|k_{so}(t-x)| \gg k_{so}/\sigma$ implies $|(t-x)| \gg 1/\sigma \approx l_c$, where
$l_c$ is the characteristic spread of the wave packet in position space. The approximate expression of $\bar{X}_{\phi}$ shows that the output signal is amplified due to 
the Unruh thermalization. However, this amplification is quite small. Since we initially assume that $k_{so}/\sigma \gg 1$, so $|k_{so}(t-x)| \gg 1$,
then $e^{- 2 \pi k_{so}|t-x|}$ must be a very small number. This can be verified in our numerical integration of Eq.(\ref{signal-general}) below.

Next, we would like to calculate the variance of the signal. Using Eq.(\ref{coefficients}) and the identity
\begin{equation}
\int \frac{d k_s}{2 \pi k_s} k_s^{i (k_d - k'_d)} = \delta(k_d - k'_d),
\end{equation}
we find
\begin{eqnarray*}
\langle 0 | \hat{b}_{k_d}\hat{b}^{\dag}_{k'_d} | 0 \rangle = \int d k_s A_{k_d k_s} A^*_{k'_d k_s} = \frac{1}{1 - e^{- 2 \pi k_d}}\delta(k_d - k'_d), \\
\langle 0 | \hat{b}^{\dag}_{k'_d}\hat{b}_{k_d} | 0 \rangle = \int d k_s B_{k_d k_s} B^*_{k'_d k_s} = \frac{e^{- 2 \pi k_d}}{1 - e^{- 2 \pi k_d}}\delta(k_d - k'_d),
\end{eqnarray*}
and therefore,
\begin{flalign*}
&\langle 0 | \{\hat{b}_S(\tau), \hat{b}^{\dag}_S(\tau') \} | 0 \rangle = \int d k_d \int d k'_d f_S(k_d,\tau) f^*_S(k'_d,\tau')&  \\
& \times \langle 0 | \{\hat{b}_{k_d},\hat{b}^{\dag}_{k'_d}\} | 0 \rangle &\\
&= \int d k_d f_S(k_d,\tau) f^*_S(k_d,\tau') \times  \frac{1+e^{- 2 \pi k_d}}{1 - e^{- 2 \pi k_d}}.&
\end{flalign*}
Taking into account $f_S(k_d,\tau)=f_L(k_d,\tau)$, we have
\begin{flalign}\label{variance-general}
&V = \beta^2 \sqrt{\frac{2}{\pi}} \frac{\sigma}{k_{so}} \int d k_d  \bigg \{ e^{-2\sigma^2 [k_d + k_{so}(t-x)]^2/k^2_{so}} & \nonumber \\
& + 2 \text{cos}[2 k_{so}(t-x)] e^{-\pi k_d}e^{-\sigma^2 [k_d + k_{so}(t-x)]^2/k^2_{so}} & \nonumber \\
& \times e^{-\sigma^2 [k_d - k_{so}(t-x)]^2/k^2_{so}} + e^{-2\pi k_d}e^{-2\sigma^2 [k_d - k_{so}(t-x)]^2/k^2_{so}} \bigg\}& \nonumber \\
&\times  \frac{1+e^{- 2 \pi k_d}}{(1 - e^{- 2 \pi k_d})^2}. &
\end{flalign}
Substituting Eq.(\ref{variance-general}) into Eq.(\ref{normalized variance}), we finally get a general expression for the normalized variance of the 
output signal. Again, in the case where $t-x<0$ and $|k_{so}(t-x)| \gg k_{so}/\sigma$, we can recover the analytic expression found in \cite{Downes13},
\begin{equation*}
\bar{V}_{\phi} \approx \frac{1+e^{- 2 \pi k_{so} |t-x|}}{1 - e^{- 2 \pi k_{so} |t-x|}}. 
\end{equation*}
However, the Unruh thermalization effect is still very small because $|k_{so}(t-x)| \gg 1$ so $\bar{V}_{\phi} \approx 1$.


\section{Horizon-straddling case}\label{horizon-straddling}
We would like to explore the horizon-straddling case where $t-x \approx 0$. The approximation made in \cite{Downes13} is no longer valid because
contributions of the second and third terms in Eqs.(\ref{signal-general}) and (\ref{variance-general}) are significant and important. Since there
is no analytic expression for the integration, we numerically integrate Eqs.(\ref{signal-general}) and (\ref{variance-general}) for various 
parameters. It turns out that in most cases $I$ and $\bar{V}_{\phi}$ are divergent if we integrate over an arbitrarily low frequency. Physically, that means if Rob's detector is strong
enough such that it can detect arbitrarily low frequency particles, then Rob will observe a large expectation value and fluctuation of the number of low frequency particles.
This is reasonable because when the wave packet straddles Rob's future horizon, most of these particles are greatly redshifted as seen by Rob, especially at late 
times when Rob's velocity approaches the speed of light. In realistic situations, Rob's detector cannot detect arbitrarily low frequency particles. 
Therefore, we introduce a low frequency cutoff $k_{\text{cut}}$ for the detector mode function. 
One might expect that the low frequency cutoff depends on the specific detector Rob carries. That is true, but we do not want to
discuss specific models of Rob's detector. We can find a natural low frequency cutoff by other considerations.

\begin{figure}[ht!]
\includegraphics[width=9.0cm]{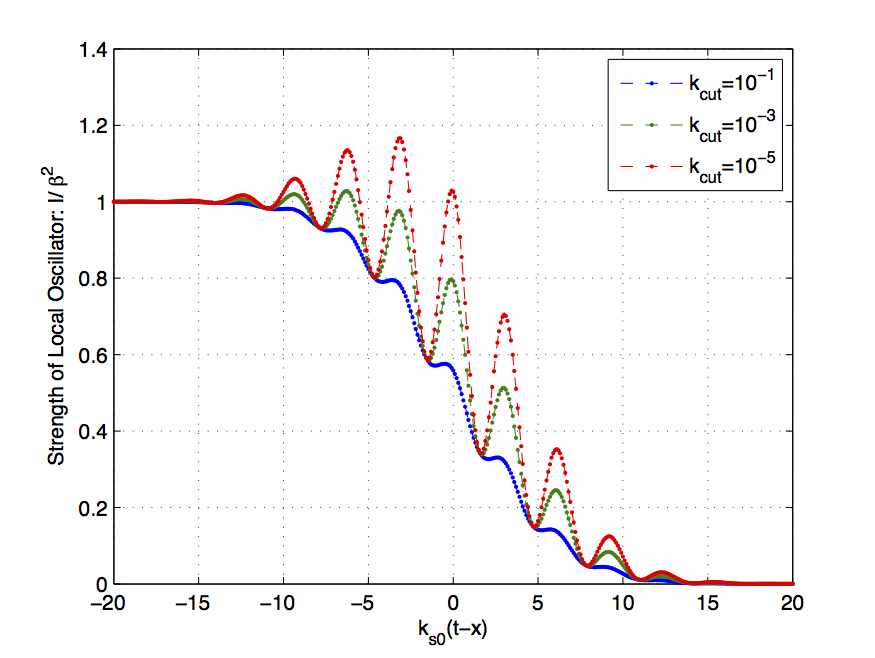}
\caption{\footnotesize (color online). Strength of local oscillator for various low frequency cutoffs: $k_{\text{cut}} = 0.00001(\text{top}), 
0.001(\text{middle}), 0.1(\text{bottom})$, $\delta = \frac{k_{so}}{\sigma} = 10$.} 
\label{signal:Fig}
\end{figure}

\begin{figure}[ht!]
\includegraphics[width=9.0cm]{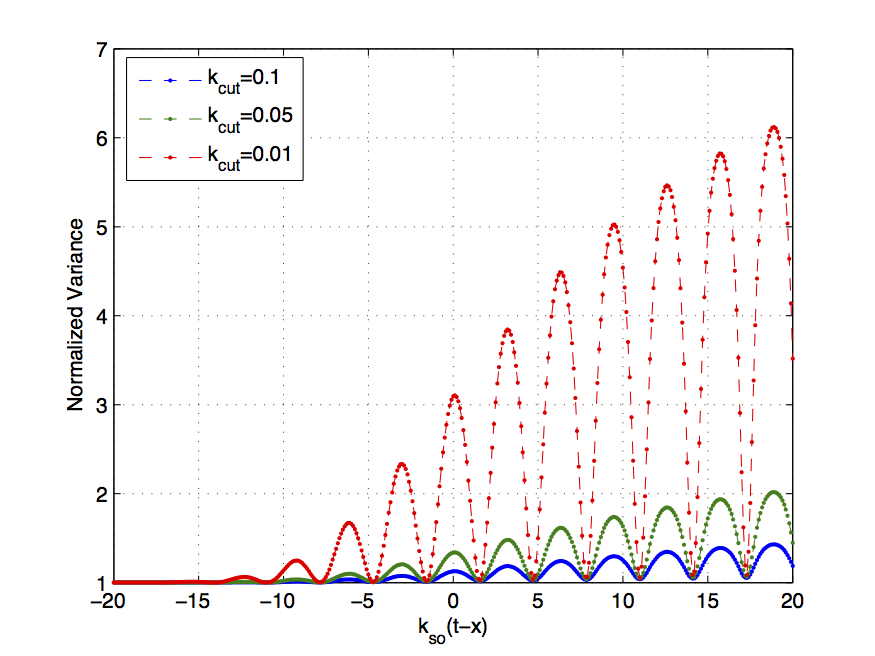}
\caption{\footnotesize (color online). Normalized variance for various low frequency cutoffs: $k_{\text{cut}} = 0.01(\text{top}), 
0.05(\text{middle}), 0.1(\text{bottom})$, $\delta = \frac{k_{so}}{\sigma} = 10$.}
\label{variance:Fig}
\end{figure}

Figs.\ref{signal:Fig} and \ref{variance:Fig} show the strength of the local oscillator and the variance of the output signal received by Rob
for various parameters. According to Eq.(\ref{normalized signal}), the strength of the local oscillator $I/\beta^2$ also characterizes the 
amplitude of the expectation value of the output signal for a given relative phase $\phi$. Thus Fig.\ref{signal:Fig} also indirectly shows the 
amplitude of the expectation value of the output signal. 
We can see that they depend on when Alice sends the signal and local oscillator if the central wave number 
$k_{so}$ is fixed. If Alice sends the signal and local oscillator early enough then $I \approx \beta^2, \bar{V}_{\phi} \approx 1$, and thus 
$\bar{X}_{\phi} \approx \alpha e^{i \phi} + \alpha^* e^{-i \phi}$. Rob sees the original coherent state signal. The Unruh 
thermalization effect is not significant, as we have argued before. If Alice sends them later so that the wave packet straddles Rob's future
horizon, the strength of the local oscillator decreases
with some characteristic oscillation, while the variance increases with similar oscillation. The Unruh thermalization becomes significant in this 
horizon-straddling case. Interestingly, if we choose lower frequency cutoff, for some specific values of $k_{so}(t-x)$ the strength of the local oscillator
and the variance remain unchanged, while for other $k_{so}(t-x)$ they increase dramatically. These particular values of $k_{so}(t-x)$
can be determined by $k_{so}(t-x) \approx (\frac{1}{2} + n)\pi, n = 0, \pm 1, \pm 2,... $, and at these points the variances are approximately one.
From Eq.(\ref{signal-general}), the local oscillator received by Rob is quite different from that sent by Alice in the horizon-straddling case.  
Since Rob still can see the wave packet at late times when his velocity approaches the speed of light, one expects that the wave packet is greatly
redshifted as seen by Rob. Therefore, Rob's effective local oscillator consists of large amounts of low frequency components, resulting in 
large expectation value and variance in the homodyne detection, implying an amplification of the original coherent state.
However, for some specific values of $k_{so}(t-x)$, the low frequency components in the local oscillator
are strongly suppressed. This can easily be verified by substituting  $k_{so}(t-x) = (\frac{1}{2} + n)\pi$ into the integrand in
Eq.(\ref{signal-general}). Consequently, the strength of the local oscillator and the variance do not significantly depend on the low
frequency cutoff for these values of $k_{so}(t-x)$. 

\begin{figure}[ht!]
\includegraphics[width=9.0cm]{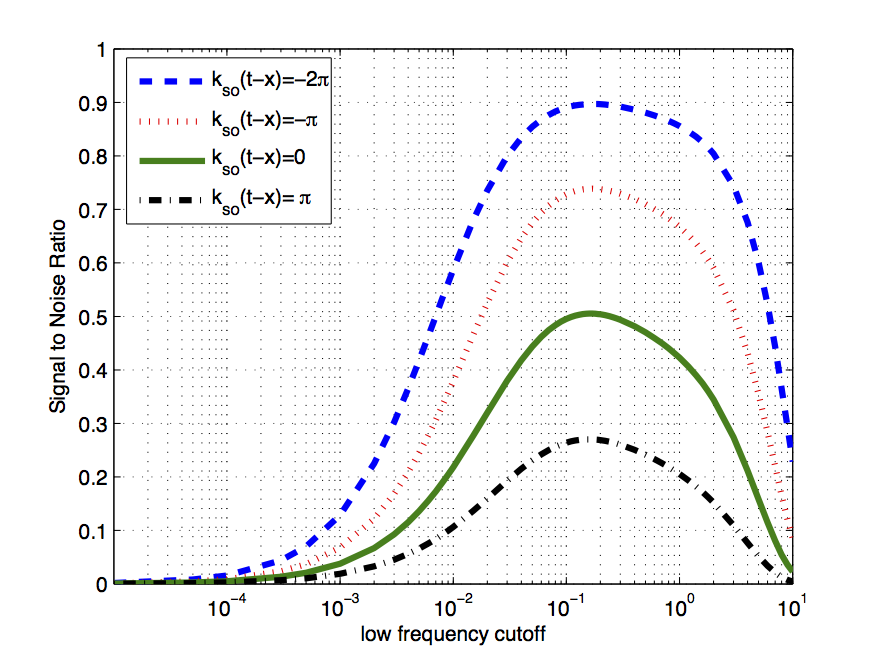}
\caption{\footnotesize (color online). Signal to noise ratio versus low frequency cutoff for $k_{so}(t-x) = n\pi$, $\delta = \frac{k_{so}}{\sigma} = 10$.
The signal to noise ratio decreases when the low frequency cutoff become smaller and larger. The low frequency cutoff that maximizes the signal to noise
ratio is between $0.1$ and $0.2$.} 
\label{SNRpeak:Fig}
\end{figure}

\begin{figure}[ht!]
\includegraphics[width=9.0cm]{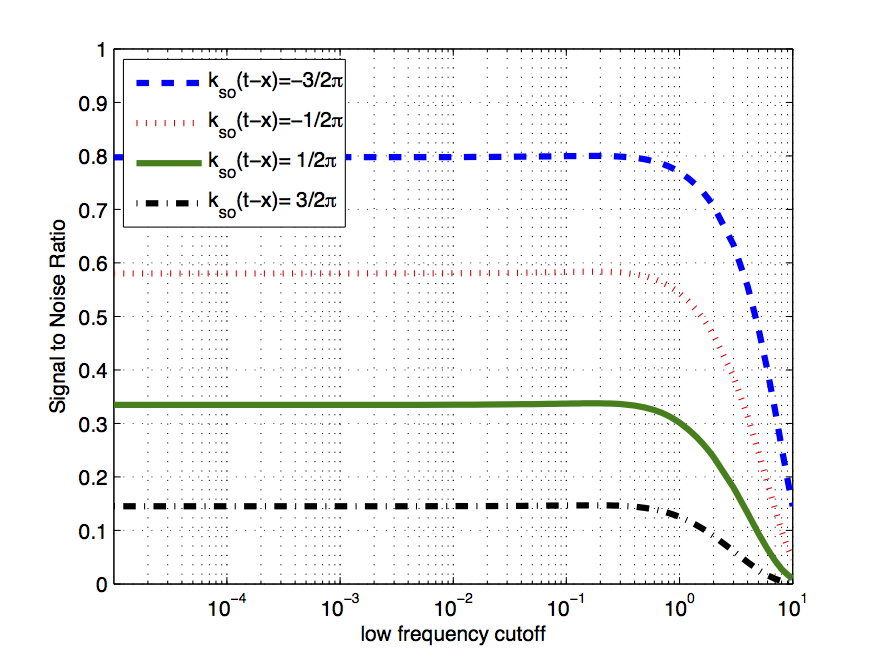}
\caption{\footnotesize (color online). Signal to noise ratio versus low frequency cutoff for $k_{so}(t-x) = (\frac{1}{2} + n)\pi$, $\delta = \frac{k_{so}}{\sigma} = 10$.
The signal to noise ratio first increases and then tends to be a constant when the low frequency cutoff becomes smaller.} 
\label{SNRtrough:Fig}
\end{figure}

Figure.\ref{SNRpeak:Fig} shows Rob's signal to noise ratio for $k_{so}(t-x) = n \pi$. These values approximately correspond to peaks of the oscillation of the 
expectation value and variance of the output signal, as shown in Figs.\ref{signal:Fig} and \ref{variance:Fig}. The signal to noise ratio decreases and goes to zero
when the low frequency cutoff becomes smaller. This is because the variance increases faster than the expectation value as the low 
frequency cutoff approaches zero. On the other side, when the low frequency cutoff becomes larger, the signal to noise ratio also decreases.
Since the variance tends to one in the large low frequency cutoff limit, this means the expectation value of the output signal decreases. There
is a maximum when the low frequency cutoff is between $0.1$ and $0.2$. The behavior of the signal to noise implies that the signal and
local oscillator Rob receives mainly contain low frequency particles. However, when $k_{so}(t-x) = (n + 1/2) \pi$ where troughs of the oscillation
of the expectation value and variance of the signal locate, the behavior of the signal 
to noise ratio is a bit different. Instead of going to zero, it tends to be constant when the low frequency cutoff is smaller than some particular value,
which is also between $0.1$ and $0.2$, as can be seen from Fig.\ref{SNRtrough:Fig}. This is closely related to the fact that for these values of 
$k_{so}(t-x)$ the low frequency components in the local oscillator are strongly suppressed. For those values of $k_{so}(t-x)$ between peaks and troughs,
the signal to noise ratio behaves more like those at the peaks, because both the expectation value and variance increase but the variance increases 
faster than the expectation value in the low frequency limit. Therefore, we can see that there exists a low frequency cutoff $k_{cm}$ which maximizes
the signal to noise ratio for various $k_{so}(t-x)$ and $k_{cm} \approx  0.15$. 
An interesting observation is that the low frequency cutoff that maximizes the signal to noise ratio is approximately corresponding to the Unruh temperature
(we employ units with $\hbar = k_B = c = 1$), 
\begin{equation}
\omega_{cm} = k_{cm} a \approx \frac{a}{2 \pi},
\end{equation}
where $a$ is the proper acceleration of Rob. In communication of classical information using quantum states, the best strategy is to have a maximal signal to noise ratio.  
Therefore, the Unruh frequency provides a natural low frequency cutoff if Alice tries to send classical information to Rob via her quantum states.  

However, if Alice wants to send quantum information to Rob, it is also important to minimize the amount of noise added such that the states remain close to 
the quantum limit. This can be quantified via the conditional variance between the input and output\cite{Ralph98},
which for this system can be defined as
\begin{equation}\label{conditioanl variance}
V_C = \bigg(1-\frac{\text{SNR}_{\text{out}}}{\text{SNR}_{\text{in}}} \bigg)V_{\text{out}} = \bigg(1-\frac{I^2}{\beta^2 V} \bigg) \bar{V},
\end{equation}
where $\text{SNR}_{\text{in}}$ represents the signal to noise ratio of input state, in our case it is the coherent state signal $| \alpha \rangle$ sent by Alice;
while $\text{SNR}_{\text{out}}$ represents the signal to noise ratio of output state, in our case it is the state received by Rob. 

\begin{figure}[ht!]
\includegraphics[width=9.0cm]{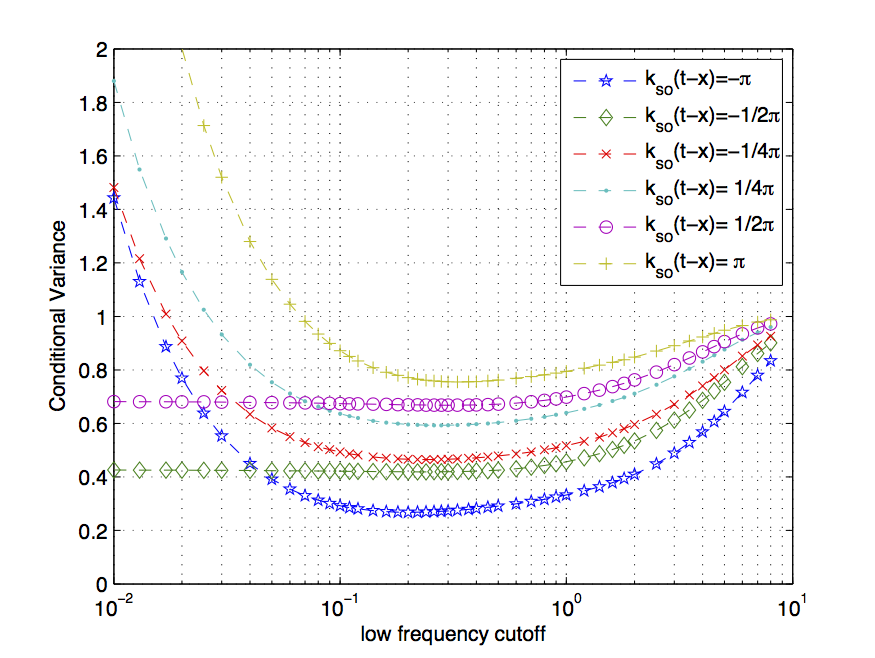}
\caption{\footnotesize (color online). Conditional variance versus low frequency cutoff, $\delta = \frac{k_{so}}{\sigma} = 10$.} 
\label{CV:Fig}
\end{figure}

Figure.\ref{CV:Fig} shows that for a given $k_{so}(t-x)\leq \delta$(horizon-straddling case), the conditional variance has a minimum. However, the location of 
the minimum slightly changes for various $k_{so}(t-x)$. Comparing with Fig.\ref{SNRpeak:Fig} one can see that locations of the minimum of the 
conditional variance do not exactly coincide with locations of the maximum of the signal to noise ratio. The former are a bit larger than the latter,
approximately ranging from $0.1$ to $0.4 $. Nevertheless, they are still in the same order of magnitude, approximately equal to the Unruh frequency. 
Therefore, we conclude that the Unruh frequency provides a natural low frequency cutoff to optimize the communication of both classical and quantum information
between an inertial partner and uniformly accelerated partner using coherent states.

\,

\,


\section{Conclusion}\label{conclusion}

We discuss quantum communication using coherent states and homodyne detection between an inertial partner and uniformly accelerated partner 
in the horizon-straddling case in which the inertial partner sends both the signal and local oscillator. 
We find that under some special conditions the accelerated partner cannot detect substantial low frequency particles regardless of his proper acceleration,
in contrast with the general viewpoint that the accelerated observer sees large amounts of low frequency particles if their acceleration is large. We also show that 
the Unruh frequency provides a natural low frequency cutoff both for quantum limited classical communication and quantum communication between the inertial partner and
uniformly accelerated partner.

\begin{acknowledgments}
We acknowledge Jason Doukas for useful discussions. This research was supported in part by the Australian
Research Council Centre of Excellence for Quantum Computation and Communication Technology (Project No. CE110001027).

\end{acknowledgments}


\end{document}